
\input harvmac

\def\frac#1#2{{#1\over#2}}

\mathchardef\ka="101A

\catcode`\@=11
\def\slash#1{\mathord{\mathpalette\c@ncel{#1}}}
\overfullrule=0pt

\def\steepslash{\c@ncel}
\def\frac#1#2{{#1\over #2}}

\def\inbar{\,\vrule height1.5ex width.4pt depth0pt}
\def\IB{\relax{\rm I\kern-.18em B}}
\def\IC{\relax\hbox{$\inbar\kern-.3em{\rm C}$}}
\def\IP{\relax{\rm I\kern-.18em P}}
\def\IR{\relax{\rm I\kern-.18em R}}
\def\IZ{\relax\ifmmode\mathchoice
{\hbox{Z\kern-.4em Z}}{\hbox{Z\kern-.4em Z}}
{\lower.9pt\hbox{Z\kern-.4em Z}}
{\lower1.2pt\hbox{Z\kern-.4em Z}}\else{Z\kern-.4em Z}\fi}

\def\brs{{\scriptscriptstyle\rm BRS}}

\def\dda{\partial_{\alpha}}
\def\ddb{\partial_{\beta}}
\catcode`\@=12

\def\draftmode{\message{ DRAFTMODE }\def\draftdate{{\rm preliminary draft:
\number\month/\number\day/\number\yearltd\ \ \hourmin}}
\headline={\hfil\draftdate}\writelabels\baselineskip=20pt plus 2pt minus 2pt
 {\count255=\time\divide\count255 by 60 \xdef\hourmin{\number\count255}
  \multiply\count255 by-60\advance\count255 by\time
  \xdef\hourmin{\hourmin:\ifnum\count255<10 0\fi\the\count255}}}

\Title{\vbox{\baselineskip12pt\hbox{IASSNS-HEP-93/66}
                \hbox{}}} {\vbox{\centerline{On the Problems
                                with Background Independence}
\vskip6pt\centerline{in String Theory}}}

\bigskip\centerline{Samson L. Shatashvili \footnote{$\dagger$}
{Research supported by NSF grant PHY92-45317. }
\footnote{$^\#$}{On leave of absence from St. Petersburg
Branch of Mathematical Institute (LOMI), Fontanka 27, St. Petersburg
191011, Russia.}}
\bigskip\centerline{School of Natural Sciences}
\centerline{Institute for Advanced Study}
\centerline{Olden Lane}
\centerline{Princeton, NJ 08540}
\bigskip\centerline{\sl Dedicated to my teacher L. D. Faddeev}
\centerline{\sl on the occasion of his 60th birthday}
\vskip .5in

The problems with background independence are discussed in the
example of open string theory. Based on the recent proposal by Witten
I calculate the String Field Theory action in
conformal perturbation theory to second order and demonstrate that the
proper treatment
of contact terms leads to nontrivial equations of motion.
I conjecture the form of the field theory action to all orders.

\Date{October, 93}

In contrast with  difficult problems in realistic 4d Field Theory models,
where the theory is defined and an explicit analytic solution is
not yet known, String Theory isn't yet even defined. In many cases what we
have is just a number of $S$-matrices for the processes when the background
is fixed by our choice of conformal field theory in
the first quantized formulation for amplitudes. Satisfactory
formulation of String Theory
would have been a formulation where we don't need to refer to any particular
classical background and these "classical backgrounds" are given by
solutions of some equations. The latter statement is very vague, because
unfortunately it is not even clear (at least to the author) what should be
the right terminology to address the question. It is believed,
by analogy with the
second quantized description of ordinary quantum field theory, that the
understanding of vacuum structure of string theory as well as the
nonperturbative character, can be achieved by developing the field theory
language to describe the target space theory. It might well be
that  the procedure that allows us to construct second quantized
field theory from Feynman sum over trajectories directly
applied to string theory is not the best way to approach the problem and some
other new ideas should be introduced. One of the most important ingredients
of any construction has to be a background independence.

In this paper I will address the question of background independent formulation
of string theory in the example of open string theory recently
suggested by Witten \ref\w{E. Witten, Phys.Rev., D46
(1992) 5446, hep-th - 9208027.}\foot{For the discussion of the same problem
in the case of closed string field theory
in a different formalizm see \ref\sz{A. Sen, B. Zwiebach, Quantum Background
Independence of Closed String Theory,
Preprint MIT-CTP-2244, TIFR-TH-93-37, hep-th/9311009.}
and references therein.}. I can't claim that at
present every point is
understood for the case of
open string; this paper should be considered as an attempt to
single out main problems and  find a correct language based on this experience.
The calculations and  observations presented in section 2, together with
final result (see below) might serve as a proper guide.
This explains the title.

I'll show that the
integrals of total derivatives do not decouple inside the correlation functions
that defines the String Field Theory action in the formalism of \w\
 due to contact term contributions. I'll explain that these contributions have
universal
character and can't be removed by change of renormalization scheme
(this statement has the same origin as the
one for gauge anomalies in field theory). This
fact leads to slight modifications of the assumptions made in original
paper \w\ and also in
\ref\me{S. Shatashvili, Phys. Lett. B311 (1993) 83, hep-th - 9303143.}.
It was shown in \me\ that under the key assumption
of decoupling of total derivatives, plus the requirement that
BRST operator on the boundary is coupling constant independent,
theory has a linear character. I'll demonstrate here
that including
the contribution of total derivatives one also has to properly define
"BRST" operator on the boundary,
 which now necessarily should depend on couplings
to satisfy the consistency condition.
I'll discuss the ambiguities related to this issue and will make a
particular choice.
In this setup I'll calculate the  field theory action in the lines of \me\
using the conformal perturbation
theory around some fixed point and demonstrate the existence of following
relation :

\eqn\final{S=
-\beta^i\frac{\partial}{{\partial}t^i}Z(t)+Z(t),}
up to the second order in coupling constant.
Here $\beta$ is the world-sheet $\beta$-function and $Z$ is a partition
function. I think that (1)
is true to all orders, but calculations beyond second order, as usual, are
very complicated.
It was discovered in \ref\ww{E. Witten, Phys. Rev., D47 (1993) 3405,
hep-th - 9210065.}
from very general arguments that
action should
have the above form with first term in \final\
given by some vector field. Thus, this vector
field is identified with $\beta$-function
in our approach. This identification is consistent
with the statement of \ww,
that zeros of vector field are
the classical equations of motion (CFT on world-sheet).\foot{The
fact that the
string field theory action on the classical equations of
motion is given by the world-sheet bosonic partition function was
previously suspected in \ref\ft{E. Fradkin and A. Tseytlin,
Phys. Lett. B 163 (1985) 123.} \ref\cal{A. Abouelsaood, C. Callan, C. Nappi and
S. Yost, Nucl. Phys. B 280 (1987) 559.}}
We will give an alternative way of explaining this
statement in the section 2.

For the reasons that we are dealing with
interacting field theory, we are forced to loose the background
independence during the calculation of the action in the
perturbation theory, so this formalism doesn't achieve the final goal;
the approach is also coordinate system dependent.
In fact, the latter makes it difficult to reconstruct the
final answer globally, once it is computed in the perturbation theory.
But, if \final\ is correct to all orders, formal background independence
is preserved. At the same time any approach to write down the expressions
for $\beta$ and $Z$ should appeal to perturbation expansion.

I'll not discuss the important issue of gauge symmetries for \final, but
as it follows from \final, all symmetries of partition function are
automatically the symmetries of action $S$. I think
that the general transformation properties
could be written using the results of Section 2 in the lines of \w.

The form of \final\ is quite general one could conjecture that
it should be true also for the
closed string theory even the analog of \w\
(the corresponding space of 2d field theories together with
background independent formulation) is not
yet known. The expression \final\
is simple, but the objects that enter, $\beta$ and
$Z$, usually are impossible to calculate in closed form. As a result this
formalism
doesn't avoid the usual technical difficulties that are present in any
other formulation of string field theory, although it is formally background
independent and contains all string modes.

\newsec{Boundary Problem and Open String Field Theory}

In the beginning of this section we first will describe the construction of
\w\ with the emphasis of places where some assumptions are made.
The idea of the construction in \w\ is based on BV formalism. Let $M$ be a
supermanifold which
is equipped with closed, nondegenerate
odd simplectic structure
$\omega$ and $U(1)$ symmetry, called ghost number $U$. This means that in
Darboux coordinates
${\psi}$, ${\theta}$ on $M$ with $\psi$ fermionic and
$\theta$ bosonic $\omega=d{\psi}d{\theta}$ and $\omega$ has ghost number 1.
In analogy with ordinary (bosonic) simplectic
manifolds one can define the Poisson Brackets, antibracket, with

\eqn\pois{\{A,B\}=\frac{{\partial}_rA}{{\partial}t^k}\omega^{kj}
\frac{{\partial}_lB}{{\partial}t^j}.}

One can show that the following two simple facts take place:

i. If $V$ is a vector field that generates the symmetries of $\omega$,
which means that $(di_V+i_Vd)\omega=0$, then there exists a function
$S$ that

\eqn\s{dS=i_V\omega.}

ii. Vector field defined by \s\ generates a symmetry of $\omega$
for any function $S$.

{}From the above immediately follows that the Poisson brackets of function $S$,
$\{S,S\}$,
defined by \s\ is annihilated by $d$ and thus the function $\{S,S\}$
is a constant

\eqn\weak{d\{S,S\}=0.}
The equation

\eqn\bv{\{S,S\}=0}
is called the BV master equation and $S$ is the action functional if it
solves the master equation. Every solution of \bv\ is automatically gauge
invariant \w,\ref\sen{A. Sen and B. Zwiebach, A note
on gauge transformations in Batalin-Vilkovisky
theory, Preprint MIT-CTP2240, TIFR-TH-93-38, hep-th/93099027.}. ;
the variation of the action under
any symplectic transformation

\eqn\gauge{{\delta}t^i=\{t^i,K\},}
generated by hamiltonian function $K$ (odd function), is given by
${\delta}S=\{S,K\}$ and for $\{S,K\}=0$,
it vanishes; trivial transformations are given with $K=\{S,{\Lambda}\}$.

In the quantization of gauge theories the action $S$ is given on subspace
of $M$ with $U=0$, $S_0$, and we have to find $S$. That is in fact what
the Faddeev-Popov procedure does in the case of Gauge Theories. In the case of
String Field Theory, the idea of \w\ was to identify the
antibracket and vector field in terms of the world-sheet theory and thus
identify $S$ as the action of the corresponding target space theory. It was
claimed in \w\ that this can be done in the case of Open String in a background
independent way. The following identifications were proposed:

\eqn\anti{\omega: \omega({\delta}O,{\delta}O)=\int_{\partial\Sigma}d{\sigma}_1
\int_{\partial\Sigma}
d{\sigma}_2<{\delta}O(\sigma_1){\delta}O(\sigma_2)>}
\eqn\vector{V: {\delta_V}O=\{Q,O\}}
where $<...>$ formally is defined through the world-sheet theory given by
path integral corresponding to the 2d action
\eqn\bad{L_a=\int_{\Sigma}d^2z(\frac{1}{8\pi}g^{\alpha\beta}{\dda}X^i{\ddb}X^j
\eta_{ij}+\frac{1}{2\pi}b^{ij}D_ic_j)+\int_{\partial\Sigma}d{\theta}V(X,b,c,t)}
Here, the first term is the closed string background
and the second term describes
an arbitrary boundary interaction,
parametrised by coupling constants $t^i$ (in general there are infinite
number of coupling constants), with the condition that
the boundary operator $V$ has the form
\eqn\deff{V=b_{-1}O,}
with $O$ being a general operator of ghost number 1. $Q$ is a BRST operator
defined by BRST current: $Q=\int_Cd{\sigma}J_{BRST}$ with contour $C$
approaching
the boundary $\partial\Sigma$ \foot
{We will not worry about generality and
assume that $\Sigma$ is just a disc.} and
$b_{-1}={\int}_Cb(v)$, $b(v)=v^ib_{ij}\epsilon^i_kdx^k$
with $v^i$ being the killing vector that generates the
rotation of disc. This world-sheet action is also equipped with an ultraviolet
cutoff a.

{}From the above identifications we have the definition of string
field theory action:
\eqn\defff{dS=\frac{1}{2}\int_0^{2\pi}d{\theta_1}d{\theta_2}<dO({\theta_1})
\{Q,O\}({\theta_2})>,}
where $d=dt^id/dt^i$ and
$<...>$ again denotes un-normalized correlation function.
Witten has shown that  \anti\ gives a closed form and it is invariant
under \vector. We  need for future use to repeat his arguments and stress
the points where some assumptions are made. \foot
{One should note that the
action defined by \defff\ differs from Zamolodchikov's $c$-function
\ref\zam{A. B. Zamolodchikov, Yad. Fiz. 46 (1987) 1819, Sov. J. Nucl. Phys.
46 (1987) 1091.},
but like a $c$-function, it has to have a local minimum
at points where world-sheet theory
is conformal invariant. Probably \defff\ could be considered as a boundary
problem version of $c$-function.}

The fact that $\omega$ defined by \anti\ is closed follows from the identity:

\eqn\iden{0=<b_{-1}(A_1(\theta_1)...A_n(\theta_n))>}
Here we use the definition of $b_{-1}$ and take two limits: first we shrink
the contour C to a point and get zero; second we take the limit when
the contour approaches the boundary $\partial\Sigma$ and get
the right hand side in \iden. Thus, in the notation $\delta_iO=
\frac{\partial}{{\partial}t^i}O$ we have:

\eqn\ddw{\eqalign{&d{\omega}(\delta_iO,\delta_jO,\delta_kO)=
\frac{\partial}{{\partial}t_k}\omega(\delta_iO,\delta_jO)-\cr
&-{\rm cyclic\, permutations}=<(b_{-1}{\delta_k}O)\delta_iO\delta_j
O>-{\rm cyclic\, perm.}=0}}
and the last step follows from \iden.

BRST invariance of \anti\ is equivalent to exactness of the right hand side in
\defff. This follows from the simple observation that because the
transformation
law of $\omega$ is $\omega^{'}=\omega +
\epsilon(i_Vd+di_V)\omega$ and we already
have shown that $d\omega=0$, what we have to show is that $di_V\omega=0$. We
have

\eqn\invar{\eqalign{d<dO\{Q,O\}>=&<(b_{-1}dO)dO\{Q,O\}>-\cr
-&<dO\{dQ,O\}>-<dO\{Q,dO\}>}.}
If we use the identity \iden\ for the first term in \invar\ and the definition
\deff\ we get:
\eqn\ident{\eqalign{&<(b_{-1}dO)dO\{Q,O\}>-<dOb_{-1}dO\{Q,O\}>+\cr
&+<(dO)^2[L_0,O]>-<(dO)^2[Q,V]>=0.}}
We are considering a deformation of the critical string, so we can drop
all  terms of the type
$<\{Q,...\}>_0$, where subscript $0$
means the expectation value in the unperturbed
theory of some number of operators, using the
 argument of the contour deformation
in the definition for $Q$,
and thus in the last term of \ident\ we can
integrate by parts in the path integral to obtain $+<\{Q,(dO)^2\}>$; the same
is true for the last term in \invar, which leads to
$+\frac{1}{2}<\{Q,(dO)^2\}>$. The first
two terms in \ident\ are equal and contribute as $2<(b_{-1}dO)dO\{Q,O\}>$.
Combining all the terms we get after some cancellations:
\eqn\invari{d<dO\{Q,O\}>=-<dO\{dQ,O\}>+\frac{1}{2}<(dO)^2[L_0,O]>=0}
and we see that $\omega$, defined by \anti\ is BRST invariant only if the
right hand side in the equation \invari\ is identically zero.

It was concluded in \w\
that \anti\ is BRST invariant because of the following two assumptions:
\eqn\a{\frac{{\partial}Q}{{\partial}t^i}=0}
\eqn\b{[L_0,O]=\int\frac{\partial}{{\partial}\theta}O(\theta)d{\theta}=0}
here, in \b, the first identity means that $L_0$ is a
generator of the rotation of circle, and the second
identity assumes that total derivatives decouple
inside the correlation functions.\foot
{I would like to thank K. K. Li and
Erik Verlinde for discussions on importance of \a\ in \w\ (see also \me\ and
\ref\kk{K. K. Li and E. Witten,
Phys. Rev. D, 48 (1993) 853, hep-th - 9303067. }) and
E. Witten, who insisted that the whole perturbation theory
should be used for proper definition of BRST commutator in \defff.}

Comment: the second correlator in \invari, the one with total derivative inside
the integral, generically is not zero and might receive the contribution from
the boundary of moduli space (position of operators on the circle are moduli).
Thus, we have to treat such terms and include their contribution.
Or,  if we want to set up such a scheme during the evaluation of
correlation functions in \defff, when \a\ is satisfied
once inside the correlator, we have to
make sure that our regularization scheme leads to decoupling of
total derivatives in the second term in \invari. The latter is a nontrivial
statement and in the next section
we are going to address this question in detail. The identity
in \invari\ should be considered as the requirement for operator $Q$; so,
$Q$, when the contour approaches the boundary $\partial\Sigma$ should depend
on the couplings according to equation \invari\ and this leads to
consistency condition on the construction. From the point of
view of conformal perturbation theory the above requirement
means that we have to use
the parallel transport of $Q$, consistent with \invari\
 when we move away from the critical point $t_{CFT}$.
It happens that only the $PSL(2,R)$ subalgebra of
Virasoro algebra is relevant, thus what one needs is to deform
this subalgebra by including the contributions of boundary term.
In the next section we will
evaluate the right hand side in \defff\ and formulate this consistency
condition
in more clear terms for the case when the boundary interaction doesn't mixes
ghosts and matter.

At the end of this section as an illustration I would like to
discuss a known example of perturbation of a conformal field
theory (closed string) by dimension one operator, where the
decoupling of total derivatives doesn't takes place and
the obstruction is a ${\beta}$-function \ref\pol{A. Polyakov,
 Unpublished, A. M. Polyakov, Gauge Fields and Strings, Harwood, 1991.}.
Similar calculations
will be performed in the next section for open
strings.

Consider some CFT perturbed by a dimension one operator $V_it^i$.
\foot{I would like to thank A. Polyakov for pointing out to
me the following example.}
We will denote the correlation functions in the perturbed
theory by $<<...>>$ and those in the unperturbed theory by $<..>$; so,
the partition function for can be written as $<<1>>$,
or $<exp(i{\int}V)>$.
One can
calculate the trace of stress-tensor in the
perturbed theory in the following way.
We start from the expectation value of the holomorphic part of stress
tensor $<<T(z,{\bar z})>>$ and use the operator expansion algebra

\eqn\ope{T(z,{\bar z})V_i(w) = \frac{1}{(z-w)^2}
 + \frac{1}{z-w}\frac{\partial}{{\partial}w}V(\omega) + ... =
\frac{\partial}{{\partial}w}(\frac{1}{z-w}V + ...)}
(here we used the fact that $V$ has dimension one in the
CFT that we are perturbing;
we are speaking about closed string) to deduce that
expectation value of the stress tensor is given by total
derivatives, integrated over the points where the operator is inserted:

\eqn\pol{\eqalign{&<<T(z,{\bar z})>>=\cr
&={\sum}_i<{\int}d^2w_i
\frac{\partial}{{\partial}w_i}
(\frac{1}{z-w_i}V(w_i)+...){\sum}_n\frac{1}{(n-1)!}({\int}V)^{n-1}>.}}
If we claim that the contribution of the
total derivative in the right hand side is zero, we will conclude that the
expectation values of stress tensor is zero; but
the latter is wrong -- we know that the following relation holds:

\eqn\rel{\frac{\partial}{{\partial}{\bar z}}<<T(z,{\bar z})>> =
\frac{\partial}{{\partial}z}
<<trT(z,{\bar z})>> = {\beta}^i\frac{\partial}{{\partial}z}<<V_i(z,{\bar z})>>}
Thus the obstruction for decoupling is the $\beta$-function,
so we have to calculate the contribution of contact terms.
 We have to use the operator expansion algebra for $V_i$ to proceed further:

\eqn\opee{V_i(w_1,{\bar w_1})V_j(w_2,{\bar w_2})=\frac{g_{ij}}{|w_1-w_2|^4}+
\frac{C^k_{ij}}{|w_1-w_2|^2}V_k(w_2)+...}
Now we see that because of the poles in \pol\ and \opee, there is nonzero
boundary contribution in the integral over $w_i$ in \pol. For this we
substitute
\opee\ in \pol. After integration over $w_i$ in \pol\ we are left with the
boundary contour integral, with small contour surrounding
each point $w_j$; denoting $w_i=w_j+{\rho}e^{i\theta}$ with small $\rho$ ,
we get:

\eqn\poll{\eqalign{<<T(z,\bar z)>>=&{\sum}_{ij}t^it^j
<{\int}d^2w_j{\int}d{\theta}{\rho}e^{-i\theta}
\frac{1}{z-w_j-{\rho}e^{i\theta}}
(\frac{g_{ij}}{\rho^4}+\cr
&+\frac{C_{ij}^k}{\rho^2}V_k(w_j)+...)\sum_n\frac{1}{(n-2)!}({\int}V)^{n-2}>.}}
If we expand the denominator in \poll\ in the powers of $\rho$
and integrate over $\theta$, we see that only
second term contributes, with final answer:

\eqn\polll{<<T(z,{\bar
z})>>=C_{ij}^kt^it^j{\int}d^2w\frac{1}{(z-w)^2}<<V_k(w)>>+...}

Thus, we obtain the
desired formula \rel\ in the second order for $\beta$-function.


\newsec{Modifications and Perturbative Calculation}

In this section we consider the situation when
ghosts and matter are decoupled. This means that the boundary interaction
$V$ is a functional purely of the matter coordinates $X$ and the operators
$O$ are just $O=cV$. The following calculations were largely already
presented in \me, but it was assumed that \a\ and \b\ are valid assumptions.
As a result the calculations have captured only the linear part of $\beta$
in \final\ and the conclusion was that the theory has a linear equation of
motion. Below I will modify the construction by using the parallel transport
of $Q$ for to satisfy the consistency condition \invari. This procedure
is generally ambiguous and I will suggest the possible criteria. I would like
to note that there has to be a relation of this procedure with the issues
discussed in \ref\zw{K. Ranganathan, H. Sonoda, B. Zwiebach,
Connection on the state space over Conformal
Field Theories, Preprint MIT-CTP-2206, hep-th - 9304053. }.

On the world-sheet we are dealing with interacting field theory and thus
only the way to perform the calculations is to use the perturbation theory.
This means that in any calculation we will loose background independence
(formally), but the goal is to get final expressions in the invariant terms
that don't appeal to a particular background. Thus, we will consider
the perturbation theory near a fixed point $t_0$, which corresponds to
some conformal field theory with stress tensor $T$ and corresponding
BRST operator $Q$. When the contour $C$ approaches the boundary of the disc
the operator

\eqn\brs{Q={\int}d{\theta}c(\theta)[T_m+\frac{1}{2}T_{gh}].}
is ambiguous, or it is better to say, it needs to be defined.
We have to understand what the correct $Q$ is and make sure that the
consistency condition is satisfied.

We could write the
left hand side of equation \defff\ in the following form:

\eqn\basic{\eqalign{dS=&\frac{1}{2}\int_0^{2\pi}d{\theta_1}
d{\theta_3}<<c(\theta_1)dV(\theta_1)
c(\theta_3)\partial_{\theta_3}c(\theta_3)V({\theta_3})>>\cr
&+\frac{1}{2}\int_0^{2\pi}d{\theta_1}d{\theta_2}d{\theta_3}<<c(\theta_1)
dV(\theta_1)c(\theta_2)c(\theta_3)
[T_m(\theta_2),V(\theta_3)]>>.}}
The ghost correlation
functions in \basic\ is easy to evaluate and it is given
by a standard formula, because we consider the case when ghosts and
matter are decoupled. For the general 3-point function we have:

\eqn\gh{<<c(\theta_1)c(\theta_2)c(\theta_3)>>=2(\sin(\theta_1-
\theta_2)-\sin(\theta_1-\theta_3)+
\sin(\theta_2-\theta_3)).}

Let us treat \basic\ term by term. The expression for the first term is simple:

\eqn\bass{\eqalign{&\frac{1}{2}\int_0^{2\pi}d{\theta_1}d{\theta_2}
(2\cos(\theta_1-\theta_2)-2)
<<dV(\theta_1)V(\theta_2)>> =\cr
&={\int}d{\theta}_1d{\theta}_2cos(\theta_1-\theta_2)<<dV(\theta_1)V(\theta_2)>>
 - d<<{\int}d{\theta}V(\theta)>>+d<<1>>.}}
If we use the notation $L_n=\int_0^{2\pi}d{\theta}e^{in\theta}T_m(\theta)$,
the second term in \basic\ can be written as a combination of three
expressions:

\eqn\basics{\eqalign{\frac{1}{2}
[i(-&<<[L_{-1},{\int}d{\theta}_2V(\theta_2)]{\int}d{\theta}_1e^{i\theta_1}
dV(\theta_1)>>-c.c.)+\cr
&+i(<<[L_{-1},{\int}d{\theta}_2e^{i\theta_2}
V(\theta_2)]{\int}d{\theta}_1dV(\theta_1)>>-c.c)-\cr-
&2{\int}d{\theta}_1d{\theta}_2\sin(\theta_1-\theta_2)<<dV(\theta_1)
[L_0,V(\theta_2)]>>].}}

One can simplify \basics\ first noting that:

\eqn\ida{\eqalign{<<{\int}d{\theta}_1e^{i\theta_1}dV(\theta_1)[L_{-1},&{\int}
d{\theta}_2V(\theta_2)]>>=\cr
&=<{\int}d{\theta}e^{i\theta}dV(\theta)
[L_{-1},exp(iL^{int})]>,}}
\eqn\idb{\eqalign{<<{\int}d{\theta}_1dV(\theta_1)[L_{-1},{\int}&
d{\theta}_2e^{i\theta_2}
V(\theta_2)]>>=\cr
&=<d({exp}(iL^{int}))[L_{-1},{\int}d{\theta}e^{i\theta}V(\theta)]>.}}

We need two more identities:

\eqn\idc{\eqalign{<&d({exp}(iL^{int}))[L_{-1},{\int}d{\theta}
e^{i\theta}V(\theta)]>=d<<[L_{-1},{\int}d{\theta}e^{i\theta}V(\theta)]>>-\cr
&-<<[L_{-1},{\int}d{\theta}e^{i\theta}dV(\theta)]>>-<<[dL_{-1},{\int}d{\theta}
e^{i\theta}V(\theta)]>>,}}
and finally for the second term in \idc

\eqn\idd{\eqalign{&<<[L_{-1},{\int}d{\theta}e^{i\theta}dV(\theta)]>>=\cr
&=-<[L_{-1},{exp}(iL^{int})]{\int}d{\theta}e^{i\theta}dV(\theta)>
+<[L_{-1},{exp}(iL^{int})]{\int}d{\theta}e^{i\theta}dV(\theta)]>.}}
Here we used the notation where $L^{int}$ is the boundary interaction term,
and $dL_{-1}=dt^i\frac{\partial}{{\partial}t^i}L_{-1}$,
assuming that $L_{-1}$ might depend on the couplings.

Now we see that the difference between \ida\ and \idb, which enters in
\basics\ is given by:

\eqn\ide{\eqalign{<<{\int}d{\theta}_1e^{i\theta_1}
&dV(\theta_1)[L_{-1},{\int}d{\theta}_2V(\theta_2)]>>-\cr
&-<<{\int}d{\theta}_1dV(\theta_1)[L_{-1},{\int}d{\theta}_2e^{i\theta_2}
V(\theta_2)]>>=\cr
&=d<<[L_{-1},{\int}d{\theta}e^{i\theta}V(\theta)]>>-
<<[dL_{-1},{\int}d{\theta}
e^{i\theta}V(\theta)]>>-\cr
&<[L_{-1},{exp}(iL^{int}){\int}d{\theta}e^{i\theta}dV(\theta)>.}}

The last term in \ide\ could be dropped because it is an expectation value of
operator $[L_{-1},...]$ in a critical theory, thus we can integrate by
parts in path integral and this term is identically zero.
So, if we combine \basic, \bass\ and \ide\ we get:

\eqn\good{\eqalign{dS=d\frac{i}{2}(<<[L_{-1},{\int}d{\theta}e^{i\theta}
V(\theta)]>> - &c.c.) + dZ -
d<<{\int}d{\theta}V(\theta)>>+\cr&  + X,}}
where we denote by one form $X$ the following expression:

\eqn\abc{\eqalign{&X={\int}d{\theta}_1d{\theta}_2
\frac{\partial}{{\partial}\theta_2}sin(\theta_1-\theta_2)
<<dV(\theta_1)V(\theta_2)>>-\cr
&-{\int}d{\theta_1}d{\theta_2}
sin(\theta_1-\theta_2)<<dV(\theta_1)[L_0-\frac
{\partial}{{\partial}\theta_2},V(\theta_2)]>>-\cr
&-\frac{i}{2}(<<[dL_{-1},{\int}d{\theta}e^{i\theta}V(\theta)]>>-c.c.).}}

Now I would like to require that under proper definition
of renormalization scheme and the generators of $SL(2,R)$ subalgebra,
the object $X$ is identically zero. One should
note that the consistency condition \invari\
requires that X is just an exact form, thus there is an ambiguity
in the definition of $Q$ and equivalently the symmetries of \anti.
Moreover it is not guaranteed that in the deformed theory
the stress tensor, that enters in the definition of $Q$, is deformed
accordingly to this requirement.
Below I will give arguments that it is indeed the case and
that the vanishing of $X$ is a natural choice.
They couldn't be considered as a rigorous proof to all orders in $t$;
they are just arguments and
most likely they can lead to such a proof.
Before I turn to this very important question let
us evaluate the first term in \abc,
 to be sure that the total derivative doesn't
make it zero. We have in the lowest order in $t$:

\eqn\cijk{\eqalign{&{\int}d\theta_1d\theta_2\frac{\partial}{{\partial}\theta_2}
sin(\theta_1-\theta_2)<<dV(\theta_1)V(\theta_2)>> = \cr
&=2dt^it^jC_{ij}^k{\int}d\theta_1\frac{{sin}a}{{{sin}\frac{1}{2}a}^{\Delta_i+
\Delta_j-\Delta_k}}<<V_k(\theta_1)>>+....=\cr
&=4dt^it^jC_{ij}^k(a)<<{\int}d{\theta}V_k(\theta)>>+...,}}
with $c_{ij}^k(a)=c_{ij}^k(a/2)^{1+\Delta_k-\Delta_i-\Delta_j}$.
Here we had used the operator expansion algebra for $V_i$,
\eqn\opeeee{V_i(\theta_1)V_j(\theta_2)=\frac{c_{ij}^k}{|sin{\frac{1}{2}}
(\theta_1-\theta_2)|^{\Delta_i+\Delta_j-\Delta_k}}V_k(\theta_1)+...,}
with $\Delta_i$ being the dimension of $V_i$ in the CFT
corresponding to the coupling constant $t_0$, and integrated over
$\theta_2$ from $\theta_1+a$ to $\theta_1+2\pi-a$. The term in $c$ with
\eqn\res{\Delta_i+\Delta_j-\Delta_k=1,}
the "resonance term", is convergent in the limit when we
remove the cutoff, and others diverge.
These divergent terms can be
removed by redefinition of couplings or the same, by a subtraction procedure
(see below), while
the constant term is universal and can't be removed.
Also, there should be a higher order
correction in couplings in \cijk.

We see that, like in the example for closed string at the end of the previous
section, total derivative in $X$ doesn't decouple and is proportional to
$\beta$-function coefficient $C_{ij}^k$; thus \b\ fails, so does \a.

Until now we had avoided the question about the transformation properties
of boundary operator. To proceed further it is necessary to know the action
of $L_{-1}, L_0$ and $L_1$ on $V$'s. What we need to
cancel the contribution of \cijk\ in $X$ is:

\eqn\com{\frac{\partial}{{\partial}t^i}{\delta}_{\epsilon}V_j=
4C_{ij}^kV_k{\partial}{\epsilon}+...}
with $\epsilon=e^{-i\theta}, 1, e^{i\theta}$.

This leads to a suggestion (that has to be verified) that the proper
deformation of the action of $SL(2,R)$ algebra is given by:

\eqn\tran{{\delta}_{\epsilon}V_i={\epsilon}{\partial}V_i+{\gamma}_i^j(t)
{\partial}{\epsilon}V_j,}
and $\gamma$ is the
matrix of anomalous dimensions, which for operators $V$ are
simply given by \zam

\eqn\ga{{\gamma}^j_i(t)={\delta}^j_i-
\frac{{\partial}{\beta}^j}{{\partial}t^i}=
\delta^j_i\Delta_i+4c_{ik}^jt^k+....,}
and ${\beta}^j$ is the $\beta$-function for operator $V_j$. This deformation
is a very natural one.\foot
{There are similarities between \tran\ and the
boundary problem analog of anomaly equation \rel, recently derived by
Zamolodchikov \ref\zamm{A. Zamolodchikov, Unpublished},
see also \ref\zammm{S. Ghoshal and A. Zamolodchikov, Boundary S-matrix
and boundary state in two-dimensional integrable quantum field theory,
Rutgers Preprint
 RU-93-20, hep-th - 9306002.}. I would like to thank
A. Zamolodchikov for sharing his insight on the problem
of boundary deformations with me.}
In fact what we need is \tran\ in the first nontrivial order in couplings
to compare with \cijk, because the latter we are able to calculate only
up to this order.
Here are the arguments in support of \tran: consider the
auotomorphizms of disc:

\eqn\aut{z'=\alpha\frac{z-z_0}{1-z_0z},}
or infinitesimally
{\eqn\autt{z'={\alpha}(z-z_0+z_0z^2)}
with $|z_0|<1$ and $|\alpha|=1$.
The requirement, that corresponding $PSL(2,R)$ subalgebra of Virasoro algebra
is not broken in perturbed theory, fixes the transformation
law \tran. In the case of closed string, the transformation
properties under constant shift is standard, and
under dilatation is controlled by the Callan-Symanzik equation, that leads
to change of classical dimension to anomalous dimension \zam;
so these two elements
are universal, and the third one is fixed by the requirement, mentioned above.
The open string version is given by \tran\ and leads to \com.
We see from this expression that they are compatible in the lowest order in
couplings and guarantee that $X$ vanishes.
What is needed to complete the proof is that one has
to show the compatibility
of \tran,
in particular regularization scheme
(note that higher order terms in $\beta$-function are scheme dependent),
with the vanishing of $X$ in \abc\ and \cijk\ to all orders..

Thus, the first term in equation \good\ defines the string field theory action

\eqn\act{S=\frac{i}{2}(<<[L_{-1},{\int}d{\theta}e^{i\theta}V(\theta)]>>-c.c.)-
<<{\int}d{\theta}V(\theta)>>+<<1>>;}
and we know the transformation of boundary operator with respect to
$SL(2,R)$ from $X=0$ or \tran.
We had derived this expression in the second order for conformal perturbation
theory. At present it is difficult to
make any rigorous statement to all orders, but I think that
\act\ should be correct up to all orders. My believe is based on
important check which is provided by the expression \idd; if
one calculates the left and right hand sides of
this identity using \tran, or \com,  he will find out that these transformation
laws are consistent with \idd. For direct proof
we need to
calculate $X$ in all orders and there are two problems involved: first,
it is difficult to perform the calculation beyond second order; second,
we have to calculate the parallel transport of Virasoro
generators and derive \tran\ in the perturbation theory
(or its modification in higher orders) and make sure
that $X$ vanishes to all orders. In fact,
there are ambiguities related to both calculations,
caused by renormalization scheme dependence beyond second order.

Our final expression \act\ can be written in the form
announced in the introduction:

\eqn\final{S=
-\beta^i\frac{\partial}{{\partial}t^i}Z(t)+Z(t).}
If we remember that structure constants in \cijk\ were cutoff dependent and
this dependence we removed by a subtraction, we might have kept it up to
\final.
The reason is that as it follows
form Poincare-Dulac Theorem\foot
{The relevance of
Poincare-Dulac theorem to $\beta$-function related
issues was stressed many times by Zamolodchikov,
see \ref\zammmm{A. Zamolodchikov, Adv. St. in Pure Math., 19 (1989) 641.}.}
about vector
fields, every vector field can be linearized by appropriate redefinition
of coordinates up to the resonant terms,
and the resonant condition is related to
the zero modes of linear part $\alpha_1, ..., \alpha_n$. The $N$-th order term
can not be removed by this redefinition if and only if there exists the integer
relation of the form:
\eqn\poin{\alpha_s={{\sum}_1^N}m_i\alpha_i}
with $(m_1,...)$ integers and $m_k \ge  0, {\sum}m_k \ge 2$
and \poin\ is called the resonance relation. The linear term for
$\beta^i$ is given
by $(1-\Delta_i)t^i$, thus the resonance condition in the second order
is the one
we had written before \res: $1-\Delta_k=1-\Delta_i+1-\Delta_j$.
They correspond to finite terms in \cijk \ and can't be
removed by coordinate transformation.
This in fact proves that the non-vanishing of total derivative term in $S$ is
universal.

The expression \final\ shows that for exactly
marginal deformations of base point
(which is a particular CFT) action $S$ is the same as partition function and
this confirms the statement of \ww.
Obviously, assuming that $\beta=0$ ($\Delta_i=1$ and $c_{ij}^k=0$ in the
perturbation theory) from the begining and going through
our calculations again
we get $dS=0$. Because any attempt to calculate string field theory action
in the present approach should use the perturbation theory we can't make
any statement about global properties of action. Also, it is difficult
to check above statement about equation of motion directly
from final expression \final\ without
going to world-sheet and using the identities
described in this Section; it would be very interesting to find such
procedure.
Last important comment related to these questions:
we couldn't compare two actions
(note that dependance on $t$'s enter through $\beta$)
if they are calculated in perturbation theory
arround different points in space of $t$'s. For this we will
need the natural paralel transport in the space of
theories that we unfortunately don't have; we only know the
deformed relations for $SL(2,R)$  generators in perturbation theory.
Thus, the result is not truly background independent even it
looks so formally.

\vskip 1cm

Acknowledgments: I would like to thank A. Zamolodchikov,
E. Verlinde and E. Witten for very useful discussions.

\listrefs
\end